\documentclass[sigconf,nonacm]{acmart}
\usepackage{booktabs}
\usepackage{array}
\usepackage{xcolor}
\usepackage{tabularx}
\usepackage{threeparttable}
\usepackage{pifont}
\usepackage{multirow}
\newcolumntype{C}[1]{>{\centering\arraybackslash}p{#1}}
\newcolumntype{P}[1]{>{\raggedright\arraybackslash}p{#1}}
\usepackage{makecell}
\AtBeginDocument{%
  }

\setcopyright{acmlicensed}
\copyrightyear{2026}
\acmYear{2026}
\acmDOI{10.1145/nnnnnnn.nnnnnnn}
\acmConference[ASE 2026]{The 41st IEEE/ACM International Conference on Automated Software Engineering}{October 12--16, 2026}{Munich, Germany}

\begin{document}
\raggedbottom

\title{AOCI: Symbolic-Semantic Indexing for Practical Repository-Scale Code Understanding with LLMs}


\author{Jinshi Liu}
\orcid{0009-0004-1207-3511}
\affiliation{%
  \institution{Xingyun Zhixue (Beijing) Technology Co., Ltd.}
  \city{Beijing}
  \country{China}}
\email{liujinshi@xingyuncl.com}

\author{Hanying Zuo}
\orcid{0009-0001-0304-037X}
\affiliation{%
  \institution{Xingyun Zhixue (Beijing) Technology Co., Ltd.}
  \city{Beijing}
  \country{China}}
\email{zuohanying@xingyuncl.com}

\author{Congyin Cao}
\orcid{0009-0003-9683-173X}
\affiliation{%
  \institution{AI Application and Innovation Lab, School of New Media, Peking University}
  \city{Beijing}
  \country{China}}
\email{19913809590@163.com}

\author{Anran Zhang}
\orcid{0009-0000-1178-8364}
\affiliation{%
  \institution{AI Application and Innovation Lab, School of New Media, Peking University}
  \city{Beijing}
  \country{China}}
\email{2401112445@stu.pku.edu.cn}

\author{Yixuan Liu}
\orcid{0009-0005-7027-0633}
\affiliation{%
  \institution{AI Application and Innovation Lab, School of New Media, Peking University}
  \city{Beijing}
  \country{China}}
\email{2401112441@stu.pku.edu.cn}

\author{Xinzhou Xie}
\authornote{Corresponding author.}
\orcid{0000-0001-9294-5885}
\affiliation{%
  \institution{AI Application and Innovation Lab, School of New Media, Peking University}
  \city{Beijing}
  \country{China}}
\email{xzxie@pku.edu.cn}

\renewcommand{\shortauthors}{Liu, et al.}

\begin{abstract}
Large language models struggle with understanding codebases beyond a 
certain scale---repositories with hundreds of thousands of lines of code. 
Existing methods---retrieval, summarization, agent exploration---each 
construct a different view at query time. The view varies between runs, and what persists is typically 
ad-hoc rather than systematic.

This paper introduces AOCI (AI-Oriented Code Indexing): a 
symbolic-semantic repository representation---a structured blueprint that an LLM can read in a single pass to gain a complete repository-level picture of the system's architecture, dependencies, and key design decisions before any task. An AOCI index consists of encoding rules followed by entries, with 
one entry per code unit (file or database table). Each entry 
pairs a symbolic tag with semantic content. The symbolic component 
provides architectural coordinates; the semantic component carries 
function, dependencies, and constraints. Together they form a 
consistent, stable representation of the entire system.

Index maintenance is incremental: when code changes, only affected 
entries are regenerated under protocol rules. The AOCI Platform 
automates this process, keeping the blueprint aligned with the code.

We evaluated AOCI on four projects across three LLMs and six context 
conditions (2,160 evaluations). AOCI outperforms all deployable baselines and ranks second only to the Oracle upper bound in overall accuracy. On 19 industrial tasks across five systems, AOCI 
produced zero final-state defects, while three mainstream agent-based 
tools introduced defects in 12 tasks and consumed 4--130$\times$ more tokens 
($p < 0.001$). The advantage grows with task complexity.
\end{abstract}



\maketitle

\section{Introduction}

Large language models (LLMs) have shown great potential in coding~\cite{chen2021evaluating,feng2020codebert,wang2021codet5,hou2024llm4se,fan2023llm4sesurvey}. But when we used them to code a larger system, with more than a hundred thousand lines of code, they often failed. This often comes down to a dilemma. To understand the whole system and decide the right place to change, the LLM needs either to read the whole codebase or to guess which files to read. The first option faces a fundamental limit: even with million-token context windows, there is too much code irrelevant to the core system logic, and that causes the LLM's attention to scatter and hurts understanding~\cite{jimenez2023swe}. The second option is more commonly adopted. LLM-based coding tools today typically use retrieval-based methods to find related files, but these methods cannot find all the related files precisely, and the result is a wide range of failures — for example, it adds the wrong tables, APIs, or functions; it assumes things that don't exist and uses fields that were never there~\cite{liu2025hallucinations,tian2024codehalu}. The real problems are more varied than this short list.

To further explore the mainstream approaches researchers have used to help LLMs understand a whole repository, we saw the dilemma firsthand. We have tried syntax-oriented extraction~\cite{gauthier2023repomap} --- this approach keeps information such as function signatures and class hierarchies, but it cannot help LLMs understand our business logic clearly. We have tried natural-language summarization~\cite{oskooei2025repository} --- it blurs the system structure and causes more defects. We have tried agent-based dynamic exploration~\cite{zhang2024autocoderover, yang2024swe, wang2025openhands, zhang2024codeagent, tao2024magis, chen2025locagent} that lets LLMs automatically fetch context and generate code on demand --- however, this comes at a high token cost, and the returns drop quickly on complex business systems~\cite{yang2024swe, ma2025alibaba}.

Looking at these three lines of work together, we noticed something they all share. As our codebase grew past a hundred thousand lines, the question we ran into was no longer ``how do we compress what fits into the next prompt'' — it was ``how does the LLM keep a complete and accurate understanding of our system, run after run.'' All three approaches we had tried did attempt some kind of system view, in their own way: retrieval gave us a directory tree of the matched files; the summarizer sometimes wrote a paragraph about how a module fits into the whole; the agent built up its own picture by clicking through files. But the output depended on what the LLM happened to generate that day — in what shape, at what level of detail. The next run on the same repository looked different. Without a fixed format that says how the system should be encoded and how an LLM should decode it back, no two views of the same codebase agreed. We needed a layer the existing methods did not provide: a stable, LLM-readable blueprint of the entire system --- its architecture, dependencies, and key design decisions --- that the LLM reads once before any task and that is updated incrementally as the code evolves. That is the layer AOCI (AI-Oriented Code Indexing) fills.

An AOCI index is a structured file that can describe an entire repository with just hundreds of lines. Beginning with the encoding rules, an AOCI index mainly consists of entries, and each entry independently represents a source file or a database table. Each entry has two parts. The first part is a multi-dimensional symbolic tag, which carries structural information. The second part is a compact semantic description, covering function, relations, and interface. Since the rules and the content sit in the same file, it is convenient for an LLM to understand the whole system in a single read. Because the entries are mutually independent, changing one file does not require rebuilding the whole index.

A protocol on its own is not enough --- for this stable layer to actually exist in production, we need to keep it consistent with the code it describes. So we built an AOCI Platform on top of the protocol. The platform takes care of index generation, consistency checking, and version management, which makes the stable layer maintainable across the lifetime of a real system. One important property: the LLM can generate the index itself, so building an index does not require manual authoring from scratch.

All of this is only worth it if AOCI holds up under real testing, so we tested it both ways. On the academic side, four projects, three frontier LLMs, six context conditions — AOCI came out best among everything we could actually deploy. On the industrial side, we ran AOCI through 19 end-to-end tasks on five industrial systems (Go backend, Node.js backend, covering education, legal, and finance). To make sure the evaluation itself was honest, we added an ablation study, brought in a separate adjudication pipeline, and added explicit mitigations for pretraining contamination on public repositories. Section~5 reports the detailed results.

The main contributions of this paper are:

\textbf{1. Protocol.} \textbf{AOCI} is a symbolic-semantic indexing protocol for LLMs. It fills a layer that prior work skipped: a stable, LLM-readable view of the whole system that does not change between runs.

\textbf{2. Platform.} On top of the protocol, \textbf{AOCI Platform} keeps the stable layer in sync with the code as the code evolves. We have been using it on our production systems.

\textbf{3. Evidence.} A multi-level evaluation shows the stable layer actually works in practice: 2{,}160 academic evaluations, 19 end-to-end industrial tasks across five real industrial systems, an ablation that isolates each encoding component, and explicit mitigations against pretraining contamination on public repositories.

\section{Background and Motivation}
\subsection{Why Scaling Context Does Not Scale Understanding}\label{subsec:signal_to_noise}

Modern LLMs support context windows of 1M or even 10M tokens~\cite{gemini2024team}, yet repository-scale code understanding has not improved accordingly. The natural assumption is that a bigger window should help --- if the model can read more files, it should understand the system better. In practice, the opposite often happens.

Hong et al.~\cite{hong2025context} documented this effect across 18 frontier models under the name \emph{Context Rot}: as input length grows, model performance degrades rather than improves. Earlier results on SWE-bench~\cite{jimenez2023swe} had already shown the same pattern --- enlarging the context budget from 13K to 50K tokens reduced Claude~2's resolution rate from 1.96\% to 1.22\%. Subsequent work confirmed that this degradation persists even when all task-relevant evidence is present in the context~\cite{du2025context, chen2501coreqa}, and that LLMs are easily distracted by irrelevant content mixed in with relevant code~\cite{shi2023large}. Agentless~\cite{xia2024agentless} further showed that structured skeleton representations outperform full-file injection at repository scale, suggesting that \emph{how} information is presented matters more than \emph{how much} is included.

The reason is not hard to see. Real-world repositories are mostly low-entropy content --- imports, boilerplate, formatting, type annotations --- that carries little architectural meaning but consumes attention. The high-entropy information that actually matters --- which module owns which responsibility, how modules depend on each other, what design decisions constrain the implementation --- is scattered across files and never stated explicitly. LLMs exhibit position-decay effects in long contexts, making distant but important information harder to use for cross-file reasoning~\cite{liu2023lost,kuratov2024babilong}, and insufficient or noisy repository context drives code hallucinations such as project-context conflicts~\cite{liu2025hallucinations,tian2024codehalu}. Recent benchmarks on cross-file reasoning~\cite{ding2023crosscodeeval,liu2024repobench,deng2024r2c2} and repository-scale tasks~\cite{yu2024codereval,li2024deveval,zhang2025swebenchlive} have made these failures measurable.

Taken together, these findings point to a problem that cannot be solved by longer context alone. What the LLM lacks is not more code, but a complete picture of the system --- its structure, its intent, and its constraints --- in a form that fits the way attention mechanisms work. Moreover, the primary audience of code-related artifacts is shifting from human developers to language models~\cite{nam2024using}, which means the representation must be designed for LLM consumption rather than human readability. This raises a concrete question: do existing software documentation and code representation schemes provide such a layer?

\subsection{The Missing File-Level Intent Layer}\label{subsec:missing_layer}

Existing software documentation mainly operates at two levels: function-level artifacts, such as comments and API descriptions, and project-level artifacts, such as design documents and architectural overviews. What is largely missing between them is a \textbf{file-level intent layer} --- a structured representation that captures each file's business role, key dependencies, importance, and position within the repository architecture.

Function-level documentation describes what a function does; project-level documentation describes what the system is for. Neither captures the information most needed for repository-scale reasoning: how files relate to each other, which ones are architecturally critical, and what design constraints govern their implementation. While function-level code summarization has been extensively studied~\cite{ahmad2020transformer,leclair2020improved}, analogous systematic effort has not extended to the file level. Oskooei et al.~\cite{oskooei2025repository} build file-, directory-, and repository-level summaries but without a fixed schema that can be maintained incrementally, while Agentless~\cite{xia2024agentless} relies on directory-tree structure rather than project documentation during file localization. In industry, tools like Claude Code recommend manually maintained project knowledge files to preserve context across sessions --- a workaround that confirms the need but lacks the structure to scale. Recent research automates repository-level documentation generation~\cite{luo2025repoagent,sun2024llms4codesum}, but the outputs target human readability rather than LLM attention.

Historically, this layer was not overlooked by accident --- it was rationally skipped. Human developers fill this gap through mental models built up over months of working with a codebase, supplemented by IDE navigation and on-demand debugging. Maintaining an explicit file-level intent index would have cost more than it saved, because the only audience was humans who could already recover this information themselves. The AI era reverses this logic on three fronts simultaneously. First, the audience shifts from humans to LLMs, which have no mental model and no cross-session memory. Second, the maintainer shifts from humans to LLMs themselves, which can generate and update index entries under a fixed protocol. Third, the payoff changes from marginal to decisive --- without this layer, LLM performance on repository-scale tasks degrades sharply, as shown in Section~\ref{subsec:signal_to_noise}.

A file-level intent layer built for LLMs must therefore satisfy constraints absent from human-oriented documentation: a fixed encoding protocol that the LLM can parse unambiguously, high-density semantic elements that convey business logic without injecting raw code, and entry-level independence so that each file's representation can be read, updated, or replaced without affecting the rest. Section~3 introduces the protocol designed on this basis.

\subsection{Capability Boundaries of Existing Repository Context Methods}

To assess existing repository-context methods, Table~\ref{tab:repo_context_comparison} compares them along six dimensions covering both technical capability and deployment constraints. The results show a clear pattern of complementarity. Syntax extraction is deterministic and incrementally maintainable but weak in business semantics. Natural-language summarization helps localization but is harder to keep synchronized with evolving code. Sparse retrieval is fast and deterministic but lacks global architectural coverage. Agent-based exploration improves localization and semantic acquisition at the cost of determinism and reproducibility. These methods therefore exhibit structural trade-offs rather than isolated weaknesses.

These trade-offs are not limited to information selection --- deciding what code enters the context. Each method also makes implicit choices about information representation: retrieval returns raw code fragments, summarization produces free-form text, agents assemble context on the fly. But none of them provides a representation that is simultaneously stable across runs, globally complete, and incrementally maintainable, as the pattern in Table~\ref{tab:repo_context_comparison} confirms. Prior work on retrieval~\cite{robertson2009probabilistic,jimenez2023swe}, iterative retrieval-augmented generation~\cite{lewis2020retrieval,zhang2023repocoder,wu2024repoformer,jiang2025longrag,bi2024iterative}, repository-structure-based composition~\cite{shrivastava2023repository,agrawal2023guiding,xia2024agentless,cheng2024dataflow,wang2025rlcoder}, and graph-based representations~\cite{ma2025alibaba,ouyang2025repograph,liu2025codegraph,liu2025codexgraph} has advanced both dimensions, yet the file-level intent layer identified in Section~\ref{subsec:missing_layer} remains unfilled. This gap motivates a protocol that jointly addresses stability, coverage, and maintainability.

\begin{table*}[t]
\centering
\caption{Comparison of Repository-Level Code Context Methods}
\label{tab:repo_context_comparison}
\renewcommand{\arraystretch}{1.0}
\setlength{\tabcolsep}{6pt}
\footnotesize
\begin{tabular}{P{2.4cm}P{3.2cm}P{2.3cm}P{1.4cm}P{2.5cm}P{1.3cm}}
\toprule
\textbf{Dimension} & \textbf{Syntax Extraction} & \textbf{NL Summary} & \textbf{Sparse Retrieval} & \textbf{Agent Exploration} & \textbf{AOCI} \\
\midrule
Representative methods
& RepoMap~\cite{gauthier2023repomap}
& Hierarchical Summary~\cite{oskooei2025repository}
& BM25~\cite{robertson2009probabilistic}
& SWE-agent~\cite{yang2024swe}, AutoCodeRover~\cite{zhang2024autocoderover}, Claude Code
& This work \\
\midrule
Business semantic preservation
& -- & + & + & + & + \\
Precise file localization
& + & ++ & + & ++ & ++ \\
Global architecture coverage
& + & + & -- & + & ++ \\
Determinism and reproducibility
& ++ & ++ & ++ & -- & ++ \\
Incremental maintenance
& ++ & -- & ++ & -- & ++ \\
Context transparency and editability
& -- & -- & -- & -- & ++ \\
\bottomrule
\end{tabular}

\vspace{1mm}
\begin{minipage}{0.96\textwidth}
\footnotesize
Note: ++ indicates full support, + indicates partial support, and -- indicates weak or no support. Sparse retrieval and agent exploration follow on-demand acquisition paradigms; therefore, their scores on the ``global architecture coverage'' dimension reflect method characteristics rather than method defects. Semantic compression in AOCI inevitably loses some implementation detail; this trade-off is quantified in Section~5.
\end{minipage}
\end{table*}

\section{Design of the AOCI Protocol}

A file-level intent layer for LLMs needs three things: a fixed encoding protocol, high-density semantic elements, and entry-level independence. AOCI is a symbolic-semantic indexing protocol that addresses all three. It couples discrete symbolic tags with continuous semantic descriptions to give an LLM a complete architectural view of a repository under a limited token budget, while keeping each entry independently maintainable.

\begin{figure}[!b]
  \centering
  \includegraphics[width=\linewidth]{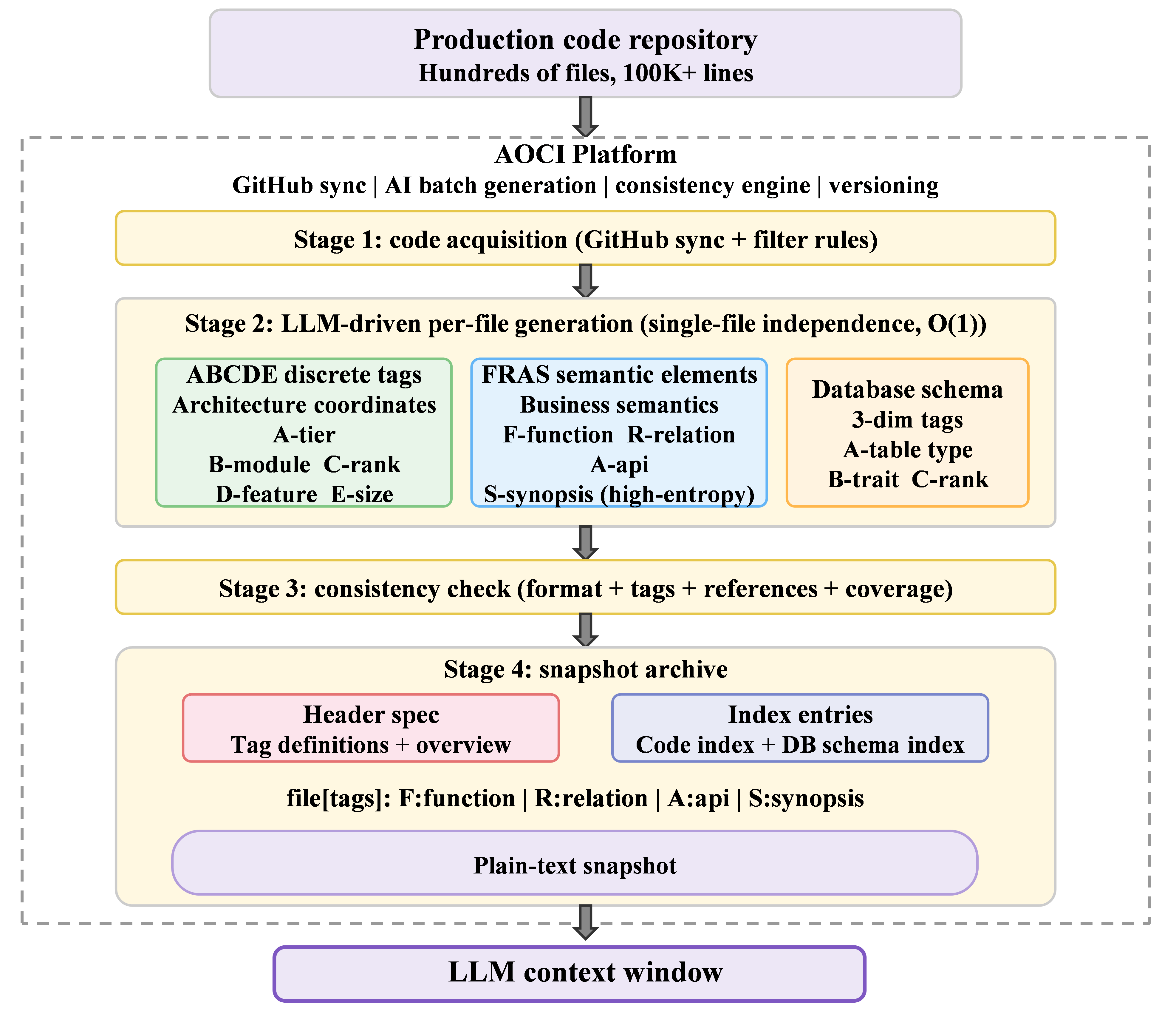}
  \caption{AOCI indexing pipeline overview.}\label{Figure1_AOCI_Architecture}
  \Description{Overview of the AOCI indexing pipeline: source repository on the left, the dual-layer encoding module in the middle converting files into discrete tags and continuous semantic descriptions, and the resulting plain-text index file on the right that is injected into the LLM context window.}
\end{figure}

\begin{figure}[t]
  \centering
  \includegraphics[width=\linewidth]{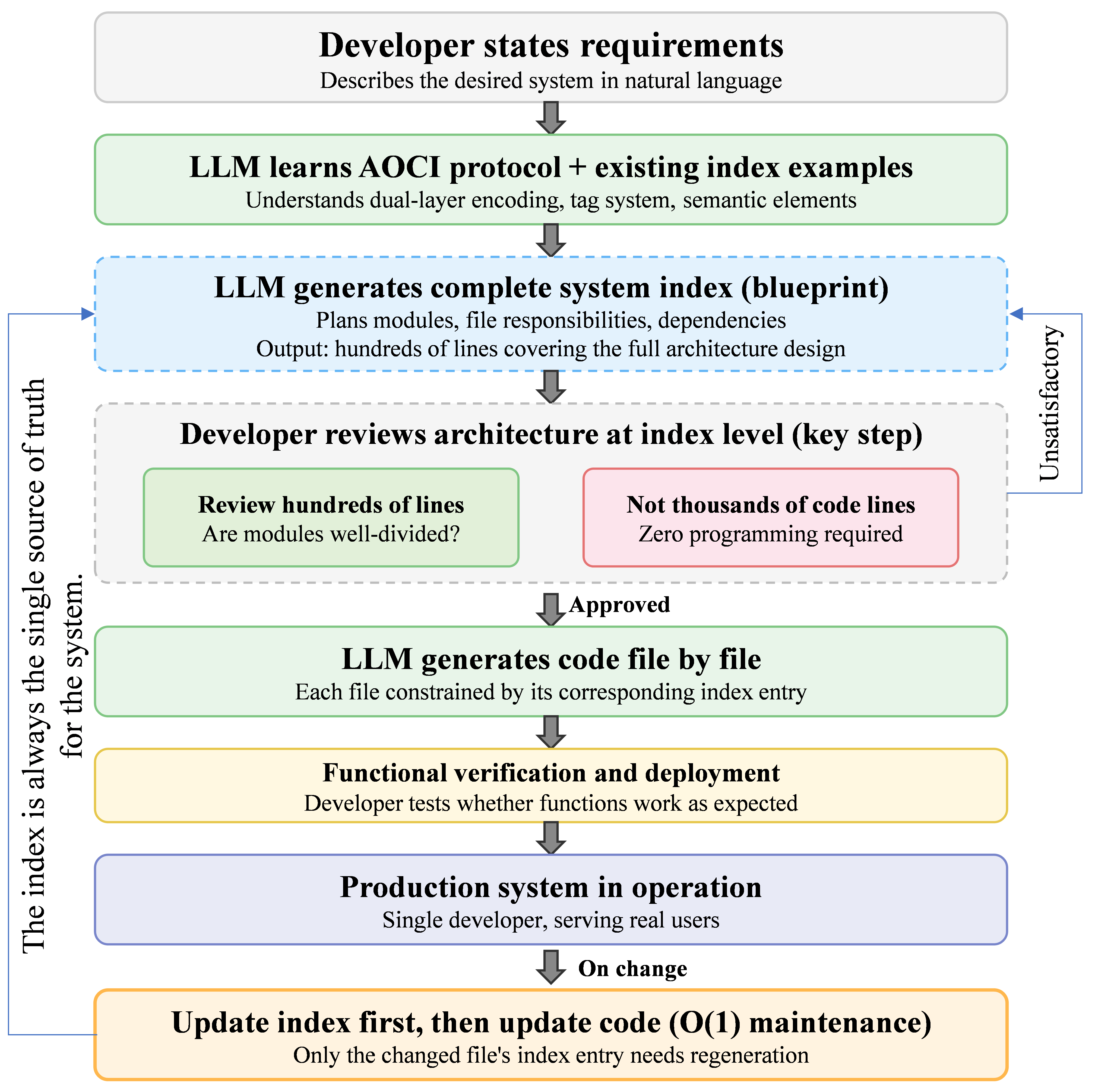}
  \caption{Index-first development workflow.}\label{Figure2_Index_First_Workflow}
  \Description{The index-first development workflow: natural-language requirements are transformed by an LLM into a complete AOCI index serving as an architectural blueprint, which is reviewed and revised at the index level before code generation.}
\end{figure}

AOCI supports two complementary modes. In \emph{code-to-index}, an LLM reads an existing repository and generates structured indexes that provide a system-level cognitive map. In \emph{index-first}, developers describe requirements in natural language, and an LLM produces a complete system index as an architectural blueprint for later implementation. Because architecture can be reviewed and revised at the index level before code generation, the review object is reduced from tens of thousands of lines of code to a few hundred lines of structured index content. Figure~\ref{Figure1_AOCI_Architecture} shows the overall AOCI architecture, and Figure~\ref{Figure2_Index_First_Workflow} illustrates the index-first workflow.

\subsection{Dual-Layer Encoding Design}

An AOCI index consists of a \textbf{header section} and a sequence of \textbf{per-file entries}. The header declares the project-level architectural overview, technology stack, and the tag taxonomy shared by all subsequent entries. Because each project develops its own business domain vocabulary, the header is defined per project. For backend entries in the AOCI Platform, for example, the A dimension dictionary contains Handler/Service/Repository/Model/Middleware/Router, the B dimension uses 30+ business module codes such as Core/\allowbreak Auth/\allowbreak Org/\allowbreak Role/Credits, and the dictionary further specifies an importance-scaled length budget where C=9 files are allocated 80 to 150 tokens and C$\leq$3 files are compressed to 20 to 40 tokens.

Each per-file entry adopts a dual-layer encoding structure consisting of a \emph{discrete tag layer} and a \emph{continuous semantic layer}, unified into a single-line plain-text format:

\begin{quote}
\texttt{filename[ABCDE-tag]: F:function | R:relations | A:API | S:synopsis}
\end{quote}

The content inside the brackets corresponds to the discrete tag layer; the content after the colon corresponds to the continuous semantic layer, with four elements separated by vertical bars. This single-line format allows a Transformer, during token-by-token processing, to first establish structural coordinates through the filename and tag anchors, and then acquire business content through the subsequent semantic elements.

The discrete tag layer consists of five orthogonal dimensions that assign each file a compact architectural coordinate. Dimension A marks the vertical architectural layer (e.g., handler, service, DAO, model, middleware, or router), enabling a layered system view. Dimension B marks the business module, providing horizontal partitioning of the domain space. Dimension C encodes file importance using a six-level non-uniform scale (\texttt{9}, \texttt{8}, \texttt{7}, \texttt{5}, \texttt{3}, \texttt{1}), which preserves finer distinctions among core files while compressing peripheral ones. Dimension D optionally records technical characteristics, such as JWT, RBAC, transactions, and encryption. Dimension E marks code scale using four levels based on line count. Concatenating these dimensions yields a compact tag string; for example, \texttt{WA9JM} denotes a medium-sized core middleware file in the authentication module with JWT-related features.

The continuous semantic layer contains four structured elements. Element F summarizes the file's business role, providing the missing file-level intent absent from traditional documentation. Element R lists major related files to capture dependencies. Element A records exposed APIs or interfaces. Element S compresses high-entropy implementation details into dense keywords, preserving design decisions such as rate limiting, fallback logic, and encryption schemes. Together, these elements form a compact semantic representation 
of a file, with core files typically under 200 tokens and 
peripheral files compressed to under 50.

The two layers are designed to complement Transformer self-attention~\cite{vaswani2017attention,xiao2024efficient}: the tag layer provides structured anchors for narrowing candidate files without full semantic scanning, while the semantic layer supplies business and design details. Because both layers are encoded within the same entry, AOCI supports localization and comprehension in a single forward pass --- delivering the stability, coverage, and entry-level independence that the file-level intent layer requires.

\subsection{Database Schema Indexing}

Code indexes primarily cover architectural information at the level of business logic, whereas complete system understanding also requires a structured representation of the data model. To this end, AOCI further extends its encoding scheme to the database layer by generating one index entry for each table and adopting a four-dimensional tag structure. Its basic form is as follows:

\begin{quote}
\small
\begin{tabular}{@{}l@{}}
\texttt{table\_name[domain-table\_type-scale\_estimate-features]:} \\
\texttt{field-level description}
\end{tabular}
\end{quote}

Here, the content inside the brackets corresponds to the table-level tags, while the remaining content consists of comma-separated field-level descriptions. The \emph{domain} dimension marks the functional area to which a table belongs, such as user, points, indexing, or auditing. The \emph{table-type} dimension distinguishes roles such as primary tables, association tables, log tables, and configuration tables. The \emph{scale-estimate} dimension categorizes tables according to their expected data volume, thereby providing the model with a prior signal about scale. The \emph{features} dimension records structural properties through multiple markers, such as JSONB fields, unique constraints, soft deletion, foreign keys, and GUID-based identification.

As an example, consider the user table in the AOCI Platform:

\begin{quote}
\small
\texttt{users[U-M-M-GUID]: user primary table,}\\
\texttt{uuid/username/email unique, password\_hash bcrypt, status,}\\
\texttt{is\_superadmin, preferences JSONB, soft delete}
\end{quote}

The tag \texttt{U-M-M-GUID} follows a dimension-by-dimension decoding scheme analogous to Section~3.1, with dashes separating the four dimensions.

\subsection{File-Level Independence and Incremental Maintenance}

Each AOCI index entry is independent, relying only on its source 
file without referencing other entries. Inter-file dependencies 
are captured by explicit filename references in the R element, 
not through embedding index content across entries. Thus, for a 
change set of $k$ modified files, only the $k$ corresponding 
entries are regenerated---$O(1)$ per changed file, independent 
of total repository size. Because R records filename-level 
relationships rather than version-pinned interfaces, entries for 
unchanged files remain valid.

Global dependencies are reconstructed at runtime through the 
Transformer's self-attention over R-element cues, avoiding a 
precomputed static graph. Tasks requiring precise call-chain 
tracing may still need the source code, as R captures 
file-level relationships rather than function-level invocations.

\subsection{Example Index Entries}

Listing~1 presents three per-file entries from the AOCI Platform project, covering three representative file categories---middleware, data access, and configuration. Complete indexes for all five industrial systems are available in the Zenodo artifact package.

\textbf{Listing 1.} Three per-file AOCI index entries from AOCI Platform

\begin{quote}
\footnotesize
\begin{tabular}{@{}l@{}}
\texttt{auth.go[WA9JM]: F:JWT authentication middleware |} \\
\texttt{R:pkg/jwt,model/user | A:- |} \\
\texttt{S:extract Bearer token from Authorization} \\
\texttt{header, parse and verify JWT, inject} \\
\texttt{user\_id and is\_superadmin into gin.Context,} \\
\texttt{expiration check, refresh logic,} \\
\texttt{API key fallback authentication,} \\
\texttt{match key\_prefix and query SHA256} \\
\\
\texttt{org\_repo.go[PO9NTM]: F:organizational data access |} \\
\texttt{R:model/org | A:- |} \\
\texttt{S:CreateWithClosure, four-step atomic} \\
\texttt{transaction, closure-table JOIN query for} \\
\texttt{GetTree, closure-table query for} \\
\texttt{GetAncestors, MoveNode subtree closure-} \\
\texttt{relation reconstruction, Delete cascading cleanup} \\
\\
\texttt{config.yaml[CC9T]: F:main configuration |} \\
\texttt{R:internal/config/config.go | A:- |} \\
\texttt{S:DB/Redis/JWT/encryption keys/rate} \\
\texttt{limiting/LLM proxy/CORS}
\end{tabular}
\end{quote}

The three tag strings (\texttt{WA9JM}, \texttt{PO9NTM}, \texttt{CC9T}) decode against the backend schema into middleware, data-access, and configuration categories. The S element preserves high-entropy design decisions
that neither syntax extraction nor natural-language
summarization captures---for instance, the API-key fallback in \texttt{auth.go}, closure-table primitives in \texttt{org\_repo.go}, and key parameter domains in \texttt{config.yaml}.

Beyond this single project, AOCI has been consistently applied across all five industrial systems with distinct technology stacks (Node.js+React, Go+Gin+Vue~3, Go+React+TypeScript), spanning backend layers (handlers, services, repositories, middleware, routers), frontend layers (pages, components, hooks, stores), configuration files, and database schemas. Full index artifacts and aggregate statistics are provided in the Zenodo artifact package and form the basis of the industrial evaluation in Section~\ref{subsec:rq2_industrial}.

\section{Evaluation Design}

AOCI is evaluated through two complementary benchmarks: an \emph{academic benchmark} and an \emph{industrial benchmark}. The former assesses representation quality by comparing six context conditions for LLM-based code understanding, while the latter evaluates end-to-end effectiveness against three mainstream agent-based tools in real production systems.

\subsection{Research Questions and Evaluation Setup}

The evaluation is organized around three research questions.

\textbf{RQ1:} \emph{How does AOCI compare with alternative context representation methods in helping LLMs understand repository architecture?}

\textbf{RQ2:} \emph{How does AOCI compare with mainstream agent-based tools on real end-to-end development tasks?}

\textbf{RQ3:} \emph{How much does each encoding component of AOCI contribute to the overall effectiveness?}

\textbf{Test Projects.}
The four academic benchmark projects follow a dual-selection strategy balancing quality control and diversity. P1 and P2 are in-house production systems whose code-quality baselines are independently established through standardized engineering verification, with details given in Table~\ref{tab:project_quality}. P3 and P4 are external GitHub projects selected for technology-stack diversity and screened for pretraining contamination, with mitigations discussed in Section~\ref{sec:threats}. The four test projects in RQ1 are P1 to P4, while the five industrial systems in RQ2 are S1 to S5. S1 and S2 correspond to P1 and P2 in industrial settings, and S3, S4, and S5 are independent systems introduced in Table~\ref{tab:industrial_benchmark}. The RQ3 ablation study uses P1.

\begin{table}[!ht]
\centering
\caption{Project Metadata and Quality Verification}
\label{tab:project_quality}
\renewcommand{\arraystretch}{1.12}
\setlength{\tabcolsep}{3pt}
\footnotesize
\begin{threeparttable}
\begin{tabularx}{\columnwidth}{
    >{\raggedright\arraybackslash}p{1.25cm}
    >{\raggedright\arraybackslash}p{1.55cm}
    >{\raggedright\arraybackslash}p{1.55cm}
    >{\raggedright\arraybackslash}p{1.35cm}
    X}
\toprule
\textbf{Project} & \textbf{Tech Stack} & \textbf{Code Size} & \textbf{Source Type} & \textbf{Quality Verification} \\
\midrule
P1: \textit{ai-xingyuncl}
& Node.js + React
& 147{,}656 LOC / 586 files
& In-house production system
& SonarQube 2A; 0 bugs / 0 vulns; Quality Gate pass; 539 tests 100\% pass. \\
P2: \textit{edu-pkuailab}
& Go + Vue 3
& Backend: 44{,}610 LOC / 200 Go files; frontend: 89 files
& In-house production system
& SonarQube 4A; 595 tests 100\% pass; 14{,}085 QPS stress; 12 sec-pen checks pass. \\
P3: \textit{dory}
& TypeScript + Next.js
& 58{,}135 LOC / 568 files
& Third-party open-source
& GitHub independent open-source project. \\
P4: \textit{koalaqa}
& Go + React + TypeScript
& 128{,}293 LOC / 707 files
& Third-party open-source
& GitHub independent open-source project. \\
\bottomrule
\end{tabularx}
\begin{tablenotes}[flushleft]
\footnotesize
\item \textbf{Note:} For P1 and P2, code-size and quality indicators are obtained from SonarQube Community Edition reports, automated tests, stress testing, and security penetration records. P3 is sourced from \texttt{https://github.com/dorylab/dory}, and P4 from \texttt{https://github.com/chaitin/KoalaQA}. LOC is measured with \texttt{cloc}, excluding dependency directories, build artifacts, and generated test files. Quad-A refers to A-grade ratings on the four core SonarQube quality dimensions (Reliability, Security, Maintainability, Coverage); ``2A'' denotes two A ratings, ``sec-pen'' denotes security penetration.
\end{tablenotes}
\end{threeparttable}
\end{table}

\textbf{Question Design.}
Each project is associated with 30 architecture-understanding questions, evenly divided into three categories by information-retrieval type. The taxonomy is adapted from the repository-level code-understanding framework of SWE-QA~\cite{peng2025swe}, merging its original four categories into three by combining \emph{Why} into \emph{How} since both require cross-file reasoning at repository scale. \emph{Where} evaluates file localization by asking for the implementation file path of a given functionality. \emph{What} evaluates dependency enumeration by asking for all files, fields, or modules related to a functionality. \emph{How} evaluates cross-module interaction understanding by asking for the full process through which a mechanism is triggered across multiple files or modules. All questions derive from actual source-code implementations, and ground-truth answers are verified with server-side \texttt{grep}, independent of any index representation. \emph{What} and \emph{How} questions must involve at least two source files, \emph{Where} questions must require cross-module reasoning beyond filename inference, and all ground-truth answers must reflect project-specific implementation decisions rather than general programming knowledge.

\textbf{Scoring.}
Scoring methods vary by category. \emph{Where} uses exact path matching. \emph{What} uses F1, the harmonic mean of precision and recall~\cite{manning2008introduction} over correctly predicted entities. \emph{How} uses LLM-as-Judge~\cite{zheng2023judging,gu2024survey} with Claude Opus 4.6 assigning an integer score from 1 to 5 based on coverage and accuracy relative to the ground truth. To reduce self-preference bias~\cite{panickssery2024llm}, the judge and evaluated models are kept in different families where possible, with CoReQA~\cite{chen2501coreqa} reporting a bias within $\pm 5.5\%$. Defect identification for the industrial benchmark follows the separate adjudication pipeline described below.

\begin{table*}[!htbp]
\centering
\caption{Context Conditions for the Academic Benchmark}
\label{tab:academic_conditions}
\renewcommand{\arraystretch}{1.2}
\setlength{\tabcolsep}{6pt}
\footnotesize
\begin{tabular}{p{0.7cm}p{2.8cm}p{3cm}p{8.5cm}}
\toprule
\textbf{Cond.} & \textbf{Method} & \textbf{Source} & \textbf{Description} \\
\midrule
A  & AOCI Index & This work & Complete AOCI structured index, injected as pure text \\
B  & RepoMap & Aider~\cite{gauthier2023repomap} & Tree-sitter-based syntax skeleton extracting signatures and class structures \\
C  & Hierarchical Summary & Oskooei et al.~\cite{oskooei2025repository} & Hierarchically aggregated file-level NL summaries \\
D  & Raw Code & Long-context baseline & Source code ranked by file size and injected into the context window up to 200K tokens \\
E1 & BM25 Retrieval & SWE-bench~\cite{jimenez2023swe} & Sparse retrieval of Top-$K$ files based on question keywords \\
E2 & Oracle Upper Bound & SWE-bench~\cite{jimenez2023swe} & Direct access to files relevant to the ground-truth answer; theoretical upper bound \\
\bottomrule
\end{tabular}

\vspace{1mm}
\begin{minipage}{0.96\textwidth}
\footnotesize
\end{minipage}
\end{table*}

\subsection{Academic Benchmark and Ablation Study}\label{subsec:academic_benchmark_design}

\textbf{Academic Benchmark.}
The academic benchmark defines six context conditions covering the major technical paths for repository-scale code understanding. Table~\ref{tab:academic_conditions} summarizes the method source and content description of each. Conditions A--D follow a \emph{global-injection} setting, where preprocessed information is injected into the context window in a single pass, while conditions E1 and E2 follow an \emph{on-demand retrieval} setting, where relevant content is selected dynamically per question. The 200K setting for Condition~D is chosen to cover 25\%--40\% of source files while staying below the threshold where performance degrades, as shown in Section~\ref{subsec:signal_to_noise}, representing a favorable upper-bound estimate for raw-code injection. Condition E2 serves as a theoretical upper bound for benchmarking the gap between AOCI and perfect information supply. All conditions use the same LLMs and question set, differing only in injected content and representation. Evaluating 30 questions with three LLMs across four projects yields 2{,}160 model calls.

\begin{table}[htbp]
\centering
\caption{Ablation Variants}
\label{tab:ablation_variants}
\renewcommand{\arraystretch}{1.15}
\setlength{\tabcolsep}{4pt}
\footnotesize
\begin{threeparttable}
\begin{tabular}{p{1.45cm}p{3.55cm}p{2.1cm}}
\toprule
\textbf{Variant} & \textbf{Description} & \textbf{Goal} \\
\midrule
Original AOCI & Complete AOCI index & Baseline \\
wo-ABCDE & Remove the entire [ABCDE] tag, keep FRAS & Role of the tag layer as a whole \\
wo-ABCD\tnote{$\dagger$} & Remove [ABCD], keep only the E (size) dimension & Signal carried by size alone \\
NL-rewrite & Rewrite full AOCI index into coherent natural-language paragraphs & Structured vs.\ free text \\
wo-R & Remove the R element (inter-file relations), keep others & Explicit dependency cues \\
wo-S & Remove the S element (high-entropy design decisions), keep others & High-entropy implementation details \\
wo-FRAS & Remove all F/R/A/S continuous semantic elements, keep only the tag & Value of the semantic layer as a whole \\
\bottomrule
\end{tabular}
\begin{tablenotes}[flushleft]
\footnotesize
\item[$\dagger$] For wo-ABCD, 8 entries (2.6\% of 307 code entries) originally without an E dimension are treated as having no valid tag and have their bracketed tag fully removed, equivalent to the wo-ABCDE treatment on those entries. This keeps the ``size-only'' semantics of wo-ABCD strictly pure. Full regeneration rules and a verification script are provided in the Zenodo artifact package.
\end{tablenotes}
\end{threeparttable}
\end{table}

\textbf{Ablation Study.}\label{subsec:ablation_study_design}
The ablation study uses project P1 with the same 30 questions and scoring protocol as the academic benchmark. Each variant changes one encoding component while keeping the others fixed. \texttt{NL-rewrite} preserves semantic content but replaces structured indexing with natural-language paragraphs, isolating the effect of representational form. Three models are evaluated across all variants, yielding 540 comparative evaluations.

\begin{table*}[t]
\centering
\caption{Industrial Benchmark Systems and Tasks}
\label{tab:industrial_benchmark}
\renewcommand{\arraystretch}{1.0}
\setlength{\tabcolsep}{4pt}
\footnotesize
\begin{tabular}{p{1.2cm}p{2.8cm}p{3.0cm}p{1cm}p{1.1cm}p{4.6cm}}
\toprule
\textbf{ID} & \textbf{System} & \textbf{Tech Stack} & \textbf{\# LoC} & \textbf{\# Tasks} & \textbf{Task Types} \\
\midrule
S1 (=P1)
& AI Practice Platform 
& Node.js + React 
& 148K 
& 3 
& Frontend changes, state persistence, frontend--backend linkage, SQL \\
S2 (=P2)
& AI Education Platform 
& Go + Gin + Vue 3 
& 82K 
& 3 
& State-machine extension, data-link synchronization, cross-module authorization \\
S3
& TE-DNA 2.0 
& Go + React 
& 56K 
& 7 
& Frontend refactoring, frontend--backend linkage, schema-sensitive operations, API contract adaptation \\
S4
& LegalMind 
& Go + React 19 + TypeScript 
& 42K 
& 3 
& Frontend components, backend cross-cutting auditing, V-shaped cross-task integration \\
S5
& AI Hedge Fund Pro 
& Go + React + TypeScript 
& 39K 
& 3 
& Data-source routing, frontend--backend linkage, complex interactive foldable panels \\
\bottomrule
\end{tabular}

\vspace{1mm}
\begin{minipage}{0.98\textwidth}
\footnotesize
\textbf{Note:} S1 and S2 are developed by the same operator. S2 underwent its own technology-stack migration to Go + Gin + Vue 3 in 3–4 days, with the AOCI index serving as the architectural anchor across the language boundary.
\end{minipage}
\end{table*}

\subsection{Industrial Benchmark}\label{subsec:industrial_benchmark_design}

The industrial benchmark compares AOCI with three mainstream agent-based tools, namely Claude Code, Cursor, and OpenCode, on end-to-end development tasks across five industrial systems built with different technology stacks. Four operators independently executed a total of 19 tasks (with per-operator loads of 6, 7, 3, and 3). Each system was developed end-to-end by a single operator, with S1--S2 sharing the same operator. Table~\ref{tab:industrial_benchmark} summarizes the system and task configurations.

All four settings use Claude Sonnet 4.6 as the underlying model. The only varying factor is the context acquisition strategy. AOCI uses precomputed structured indexes, whereas the three agent-based tools acquire context through runtime dynamic exploration under their default tool configurations. Development tasks are selected from each system's backlog to cover a range of complexities. Each task is completed by the same operator under all four settings, and comparisons are based on modified files, defects, token consumption, and completion time.

All four operators are members of the author team. To mitigate evaluation bias, defect identification across all 19 tasks is performed by a separate adjudication pipeline using Claude Opus 4.6~\cite{chiang2023can}, preloaded with the full system index, which inspects the final-state code of each setting in separate sessions without participating in task execution. Beyond per-task results reported in Section~\ref{subsec:rq2_industrial}, S1 and S2 have maintained SonarQube Quad-A ratings throughout eight months of continuous development, with 0 bugs, 0 vulnerabilities, and a cumulative 1{,}134 automated test cases passed. Systems S3, S4, and S5 also pass SonarQube Quad-A verification 
on dedicated audit runs, with partial reports provided in the 
Zenodo artifact package. These indicators reflect cumulative 
system health across all development activities and are not 
claims about any single task output.

\section{Experimental Results and Analysis}

\subsection{RQ1: Comparison Between AOCI and Alternative Context Representations}

Table~\ref{tab:cross_project_mean} summarizes the cross-model overall scores across the four projects under all six conditions. Each Finding below focuses on patterns that hold consistently across all three models. Per-model breakdowns are provided in the Zenodo artifact package.

\begin{table*}[t]
\centering
\caption{Cross-Model Mean Overall Scores Across Four Projects}
\label{tab:cross_project_mean}
\renewcommand{\arraystretch}{1.0}
\setlength{\tabcolsep}{6pt}
\footnotesize
\begin{tabular}{lrrrrrr}
\toprule
\textbf{Project} & \textbf{D* (Raw 200K)} & \textbf{E1* (BM25)} & \textbf{B* (RepoMap)} & \textbf{C* (NL Summary)} & \textbf{E2* (Oracle)} & \textbf{A* (AOCI)} \\
\midrule
P1 \texttt{ai-xingyuncl} & 26.98 & 36.69 & 39.00 & 35.05 & 92.64 & \textbf{71.07} \\
P2 \texttt{edu-pkuailab} & 45.98 & 42.03 & 38.27 & 51.55 & 91.32 & \textbf{70.11} \\
P3 \texttt{dory}         & 37.15 & 23.68 & 31.78 & 30.51 & 95.85 & \textbf{53.53} \\
P4 \texttt{koalaqa}      & 41.71 & 35.30 & 30.54 & 25.42 & 80.45 & \textbf{65.26} \\
\midrule
\textbf{Avg.}                 & \textbf{37.95} & \textbf{34.42} & \textbf{34.90} & \textbf{35.63} & \textbf{90.06} & \textbf{65.00} \\
\bottomrule
\end{tabular}

\vspace{1mm}
\begin{minipage}{0.98\textwidth}
\footnotesize
\textbf{Note:} The overall score is obtained by first averaging the model-specific scores over the three dimensions (\emph{Where}, \emph{What}, and \emph{How}) and then averaging across models. The project-level means reported in the main text are computed from the original server-side experimental records; values shown in this table are rounded to two decimal places. Per-dimension and per-model breakdowns are provided in the Zenodo artifact package. The A* column and the Avg. row are emphasized for comparison.
\end{minipage}
\end{table*}

\textbf{Finding A. \emph{Where} performance is close to the Oracle upper bound.}
On file-localization tasks, AOCI achieves a cross-project mean of 97.67\%, only 0.66 percentage points below Oracle E2 (98.33\%) and over 31 points above every other deployable baseline. This means AOCI's structured representation already contains enough information for the LLM to find the right file --- the problem the file-level intent layer in Section~\ref{subsec:missing_layer} was designed to solve.

\textbf{Finding B. The advantage on \emph{What} narrows, and a reversal appears on P3 \texttt{dory}.}
On dependency-enumeration tasks, AOCI achieves a cross-project mean F1 of 45.65\%, exceeding Condition~D by only 7.24 percentage points, a substantially smaller margin than on \emph{Where}. On P3 \texttt{dory}, Condition~D outperforms AOCI by 10.52 points consistently across all three models. P3 is an AI-driven SQL workspace whose \emph{What} questions frequently require verbatim identifier listing (e.g., SQL-field enumeration), which AOCI's semantic compression aggregates rather than preserves verbatim. This marks where the trade-off sits: cross-file dependencies are well captured, but per-identifier enumeration is lossy.

\textbf{Finding C. AOCI achieves a stronger trade-off between accuracy and token cost.}
AOCI achieves 65.00\% \emph{Overall} with 1.48M tokens, ranking second only to Oracle and consistently outperforming every deployable baseline. Relative to raw-code injection~D, AOCI attains 1.71$\times$ the score at 20.25\% of the token budget. The remaining gap to Oracle concentrates on \emph{What} and \emph{How}, consistent with Findings A and B. Unlike the other conditions, this accuracy is deterministic --- the same index produces the same result on every run.


\begin{table*}[t]
\centering
\caption{Overall Comparison Across 19 Industrial Tasks}
\label{tab:industrial_overall}
\renewcommand{\arraystretch}{1.0}
\setlength{\tabcolsep}{5pt}
\footnotesize
\begin{tabular}{lcccccccc}
\toprule
\textbf{Method} & \textbf{Total Tokens} & \textbf{Total Time} & \textbf{Total Turns} & \textbf{Functional} & \textbf{Functional} & \textbf{Architecture} & \textbf{Code} & \textbf{Cost} \\
 & \textbf{(M)} & \textbf{(S1--S2)} & \textbf{(S1--S2)} & \textbf{Bugs} & \textbf{Omissions} & \textbf{Violations} & \textbf{Residuals} & \textbf{(USD)} \\
\midrule
AOCI         & 4.29  & 15 min 50 s & 13  & 0 & 0 & 0 & 0 & \$$\sim$9   \\
Claude Code  & 41.49 & 21 min 46 s & 162 & 8 & 4 & 1 & 0 & \$$\sim$133 \\
Cursor       & 50.77 & 25 min 50 s & 205 & 9 & 0 & 3 & 0 & \$$\sim$164 \\
OpenCode     & 47.38 & 27 min 40 s & 199 & 6 & 2 & 1 & 5 & \$$\sim$152 \\
\bottomrule
\end{tabular}

\vspace{1mm}
\begin{minipage}{0.98\textwidth}
\footnotesize
\textbf{Note:}
Total Tokens are reported in millions and aggregated over all 19 tasks.
Total Time and Total Turns are aggregated over the 6 tasks from S1--S2; per-task records for all 19 tasks are provided in the Zenodo artifact package.
Token usage for AOCI is estimated from input text length, since the web-based LLM interface does not expose exact token counts. The three agent-based tools report exact API call logs.
For AOCI, Total Turns counts user--model dialogue turns. For agent-based tools, it counts tool invocation rounds.
Defects across all 19 tasks are identified by a separate adjudication pipeline using Claude Opus 4.6, preloaded with the full system index, which inspects the final-state code of each setting in separate sessions without participating in task execution. The complete defect list is provided in the Zenodo artifact package.
Cost is converted using the public pricing of Sonnet 4.6.
\end{minipage}
\end{table*}

\subsection{RQ2: Industrial Benchmark Evaluation}\label{subsec:rq2_industrial}

Table~\ref{tab:industrial_overall} summarizes the 19-task comparison. All tools use Claude Sonnet 4.6 with \texttt{temperature = 0.1}, and each task is executed once. AOCI introduces zero final-state defects across all 19 tasks. The three agent-based tools introduce defects in 12 tasks, totaling 39 defects. A Fisher's exact test on task-level defect occurrence yields \(p < 0.001\).

\textbf{Finding A. The 39 agent-introduced defects are not random --- they cluster into recurring patterns that reproduce across technology stacks.}
We observed four categories. (1)~\emph{Multi-path query omission}: a data model is accessed through multiple independent code paths, and the agent modifies some but misses others. In S1 (Node.js), all three agents found only 2 of 4~SQL queries that needed a new field. In S2 (Go), two agents missed the same field across 4~\texttt{SELECT} statements --- and Go's static type system did not catch it, because the missing field silently takes a zero value. The same bug pattern, independently reproduced across two languages. (2)~\emph{Cross-module refresh omission}: in S1, agents modified a data source but did not propagate the update to dependent views, so users saw stale counts until they manually refreshed. (3)~\emph{Interface-contract guessing}: in S3, two agents independently defined API parameters as \texttt{string[]} when the backend expected an object array --- the permission save silently failed. (4)~\emph{Schema incompatibility}: in S3, one agent produced \texttt{INSERT} statements referencing nonexistent columns (deployment crash), and another created a parallel table (data split). These are not bugs in any particular tool. They are consequences of exploring a repository at runtime without a global map. AOCI avoids them because the map already exists before the task begins.

\textbf{Finding B. The token-cost advantage grows with task complexity.}
Per-task token ratios range from 4$\times$ to 130$\times$. On S3's seven tasks spanning four complexity levels, the ratio grows from 9.4$\times$ on simple single-file changes to 11.1$\times$ on medium tasks, 30.7$\times$ on cross-module tasks, and 64$\times$ on schema-sensitive operations. S4 shows the same pattern: 15$\times$ on a single-file task, rising to 129$\times$ on a task that required understanding two prior tasks' outputs simultaneously. The reason is straightforward: agent exploration cost grows with the number of files and dependencies the task touches, while AOCI's index cost is fixed.

\textbf{Finding C. Global context prevents architectural debt and cross-task regressions, not only functional bugs.}
The agent tools did not just introduce bugs --- they introduced two additional classes of problems that a global view would have prevented. First, \emph{architectural debt}: in S2, one agent added a semantically redundant database column because it misread the design intent of an existing field, introducing a permanent schema inconsistency --- we confirmed the column was absent from the production database. In S3, two agents produced schema-incompatible code for an existing permissions table: one caused a deployment crash, the other created a parallel table with split data. Second, \emph{cross-task regression}: in S4, two agents modified a central routing file and silently dropped all parameter changes made in a prior task, reverting seven handler constructor signatures. AOCI avoids these because its index records field-level schema, design intent, and the results of prior tasks before code generation begins. S2 also serves as a cross-stack portability case: it was reimplemented from Node.js to Go in 3--4~days, with the AOCI index as the architectural anchor across the language boundary. Among the five systems, S1, S2, and S3 are in production, with S2 serving over 100{,}000 registered users.


\subsection{RQ3: Ablation Analysis of AOCI Encoding Components}\label{subsec:rq3_ablation}

This subsection conducts a component-wise ablation on project P1 \texttt{ai-xingyuncl}, selected for its mid-range complexity. Each variant is evaluated on the same 30 questions by three frontier models independently, yielding 540 primary calls and 180 \emph{How}-question judging calls. Two-sided Wilcoxon signed-rank tests with Bonferroni correction are applied across variants; Table~\ref{tab:ablation_p1_mean} reports the cross-model means and test results.

\begin{table*}[t]
\centering
\caption{Comparison of Seven Ablation Variants on Cross-Model Means for Project P1}
\label{tab:ablation_p1_mean}
\renewcommand{\arraystretch}{1.2}
\setlength{\tabcolsep}{10pt}
\footnotesize
\begin{tabular}{lrrrrrrr}
\toprule
\textbf{Variant} & \textbf{Where (\%)} & \textbf{What (\%)} & \textbf{How (\%)} & \textbf{Overall (\%)} & \textbf{$\Delta$Overall} & \textbf{$p$ (Wilcoxon)} & \textbf{Per-Question Tokens} \\
\midrule
Original AOCI & 100.00 & 45.88 & 74.67 & 73.52 & Reference & Reference & 45,949 \\
wo-ABCDE      & 100.00 & 44.00 & 77.33 & 73.78 & +0.26  & 0.500 n.s.   & 44,290 \\
wo-ABCD       & 96.67  & 42.04 & 73.33 & 70.68 & -2.84  & 0.073 n.s.   & 44,408 \\
NL-rewrite    & 93.33  & 39.87 & 70.67 & 67.96 & -5.56  & 0.026 $\dagger$      & 60,289 \\
wo-R          & 100.00 & 25.35 & 72.00 & 65.78 & -7.74  & 0.002 **     & 42,479 \\
wo-S          & 92.00  & 27.03 & 41.33 & 53.45 & -20.07 & $<0.001$ *** & 22,653 \\
wo-FRAS       & 68.67  & 8.97  & 29.33 & 35.66 & -37.86 & $<0.001$ *** & 12,299 \\
\bottomrule
\end{tabular}

\vspace{1mm}
\begin{minipage}{0.98\textwidth}
\footnotesize
\textbf{Note:}
\emph{Overall} is obtained by averaging the cross-model means over the three dimensions \emph{Where}, \emph{What}, and \emph{How}.
$\Delta$\emph{Overall} denotes the difference relative to the original AOCI condition.
The \emph{$p$ (Wilcoxon)} column reports the two-sided paired test result after combining the three models (\(N=90\)).
Significance symbols are defined as follows: *** \(p<0.001\), ** \(p<0.01\), * \(p<0.05\), and \(\dagger\) not passing the Bonferroni-corrected threshold \(\alpha=0.00833\).
The \emph{Per-Question Tokens} column reports the cross-model mean number of prompt tokens and is used to compare compression efficiency.
Exact per-model results are provided in the Zenodo artifact package.
The variants correspond to different component removals: \texttt{wo-ABCDE} and \texttt{wo-ABCD} test the discrete tag layer; \texttt{wo-R}, \texttt{wo-S}, and \texttt{wo-FRAS} test the continuous semantic layer; \texttt{NL-rewrite} tests the structured representational form.
\end{minipage}
\end{table*}

\textbf{Finding A. The S element carries the most critical information in the AOCI index.}
Removing S reduces \emph{Overall} from 73.52\% to 53.45\% (\(\Delta=-20.07\), Wilcoxon \(p<0.001\)), with the \emph{How} score dropping from 74.67\% to 41.33\% --- the largest single-component degradation. S is where the high-entropy design decisions live: rate-limiting strategies, fallback mechanisms, encryption schemes, transactional procedures. These are exactly the details that syntax extraction and natural-language summarization do not preserve (Table~\ref{tab:repo_context_comparison}). Without S, the remaining structural cues are not enough for implementation reasoning. Further removing F, R, and A (\texttt{wo-FRAS}) lowers \emph{Overall} to 35.66\%, close to Condition~D on P1 (26.98\%, Table~\ref{tab:cross_project_mean}).

\textbf{Finding B. Structured encoding is more token-efficient than equivalent natural-language prose.}
The \texttt{NL-rewrite} variant converts the full AOCI index into coherent paragraphs with 100\% identifier retention. \emph{Overall} drops from 73.52\% to 67.96\% (\(\Delta=-5.56\), Wilcoxon \(p=0.026\), not passing Bonferroni threshold), while tokens per question increase by 31.2\%. The natural-language form scores lower on all three dimensions at higher token cost, though the rewritten text is 49.8\% longer, making token overhead a confound.

\textbf{Finding C. The R element has a strong, dimension-specific effect on dependency enumeration.}
Removing R reduces \emph{Overall} from 73.52\% to 65.78\% (Wilcoxon \(p = 0.002\), passing the Bonferroni threshold). The effect concentrates on \emph{What}, which drops from 45.88\% to 25.35\%, while \emph{Where} stays at 100.00\% and \emph{How} declines only marginally.

\textbf{Finding D. The discrete tag layer does not yield a uniform aggregate gain, but exhibits model-family heterogeneity.}
Removing the ABCDE tag layer leaves \emph{Overall} nearly unchanged from 73.52\% to 73.78\% with Wilcoxon \(p = 0.500\), but the aggregate masks directional differences across models. Per-model tests show a positive effect on Sonnet, a negative effect on GPT, and no significant effect on Gemini (\(p = 0.016, 0.010, 0.965\)). The QA benchmark measures architectural understanding through Where/What/How questions, but the tag layer is designed for a different set of functions: architectural-layer navigation, importance-based prioritization, and incremental maintenance routing. These functions are directly exercised in end-to-end development (Section~\ref{subsec:rq2_industrial}), where the full AOCI index --- tags included --- produced zero defects across 19 tasks.

\section{Discussion}

The experimental results show a consistent pattern across both benchmarks. On architectural understanding, AOCI reaches near-Oracle file localization but falls short on verbatim dependency enumeration, where raw code retains literal identifiers that AOCI's semantic compression compresses into higher-level descriptions. The ablation confirms that most of AOCI's value comes from the FRAS semantic layer --- especially S, which carries the high-entropy design decisions that no other representation preserves.

The industrial benchmark reveals something different from what the academic numbers show. All three agent-based tools produced similar kinds of errors across different systems~\cite{he2024llm} --- missing cross-module side effects, silent regressions in files they did not visit, residual code from abandoned attempts. These are not bugs in any particular tool; they are consequences of exploring a repository at runtime without a global map. AOCI avoids this class of errors because the map already exists before the task begins. The advantage grows with task complexity: on simple single-file changes the token savings are modest, but on schema-sensitive and cross-module tasks the ratio exceeds 100$\times$.

Three practical lessons emerged from eight months of deployment. First, a stable repository view matters more than a larger context window --- the same index, reused across dozens of tasks, eliminated the run-to-run variance that agent tools exhibited. Second, AOCI and agent-based tools work at different layers rather than competing head-to-head; a natural next step is to use AOCI as the global context layer beneath an agent pipeline. Third, platform support for incremental maintenance is not optional --- without automated generation, consistency checking, and version management, index--code divergence would erode the stability that makes the whole approach work.

These results support the premise in Section~\ref{subsec:missing_layer}: the file-level intent layer that was rationally absent from human-oriented documentation becomes a prerequisite once the audience shifts to LLMs. AOCI's contribution is not a better retrieval algorithm or a smarter agent, but the layer itself --- stable, versioned, and designed for how LLMs actually read code.

\section{Threats to Validity}\label{sec:threats}

The LLM-as-Judge scoring of \emph{How} questions uses Claude Opus 4.6, which belongs to the same family as Claude Sonnet 4.6 and may introduce self-preference bias~\cite{panickssery2024llm}. To mitigate this, the Opus scores are aligned with independent human scoring on all 10 \emph{How} questions of P1 under Condition~A, with 8 of 10 showing exact agreement and all 10 within one point, yielding a Spearman correlation of \(\rho=0.90\) at \(p<0.001\). While absolute \emph{How} scores may be influenced by bias, the relative rankings and significance conclusions are unlikely to be materially affected.

Non-blind execution in the industrial benchmark is another threat: the same operators who developed the systems also executed the four-setting comparison. To mitigate this, defect identification is delegated to a separate adjudication pipeline using Claude Opus 4.6, as described in Section~\ref{subsec:industrial_benchmark_design}. The main experiments use single-sample evaluations with \texttt{temperature = 0.1}; most conditions exhibit substantial score differences, reducing the risk that single-sample variance affects comparative conclusions.

The evaluated projects concentrate on Web technology stacks --- Node.js, Go, and TypeScript --- and frontend--backend scenarios, which raises generalizability concerns. This is partially addressed by applying AOCI to OpenClaw, a large open-source TypeScript codebase, with pull requests generated under AOCI-guided development submitted upstream, passing CI, and receiving maintainer review. Broader practitioner replication and non-Web technology stacks remain future work.

Residual pretraining contamination in public repositories~\cite{sainz2023nlp} is reduced through pre-screening and post-hoc checks, but a more thorough exclusion requires non-public closed-source test sets. AOCI's token count is estimated from dialogue structure whereas agent-based tools use exact API logs, introducing measurement heterogeneity; the large cross-tool token ratios of 4$\times$ to 130$\times$ make comparative conclusions robust to this difference. File-level indexing granularity may be insufficient when a single file contains methods with divergent behaviors, suggesting method-level or finer-grained indexing as a future extension.

\section{Conclusion}

This paper proposes AOCI, a symbolic-semantic indexing protocol that fills a layer largely absent from existing software documentation: a stable, LLM-readable representation of file-level intent across an entire repository. Across four projects, three frontier models, and six context conditions, AOCI ranks consistently first in overall accuracy among all deployable non-oracle methods, with its file-localization accuracy reaching 97.67\%, only 0.66 percentage points below the Oracle upper bound. In the industrial benchmark, AOCI leaves no final-state defects across 19 end-to-end development tasks while consuming substantially fewer tokens than mainstream agent-based tools. The ablation study shows that most of AOCI's informational value comes from the continuous semantic layer, with the S element carrying the largest share, while the discrete tag layer serves architectural navigation and risk classification functions that fall outside the QA benchmark's measurement scope but are exercised in end-to-end development. These results support a broader observation: what LLMs lack for repository-scale understanding is not more code or longer context, but a complete and stable picture of the system --- and structured symbolic-semantic indexing provides a practical way to supply it. Main limitations include the concentration on Web technology stacks and the single-project ablation, with further details discussed in Section~\ref{sec:threats}. Future directions include hybrid designs that combine AOCI's precomputed global view with runtime agent exploration, finer-grained indexing below the file level, and controlled closed-source test sets to further mitigate pretraining contamination.

\section*{Data Availability Statement}

The artifact package is available at \url{https://doi.org/10.5281/zenodo.19677251}. It includes: (i) experimental data and scoring scripts needed to reproduce all reported results; (ii) complete AOCI index snapshots for all five industrial systems, comprising approximately 1{,}276 entries compressing over 370K LOC across distinct technology stacks; and (iii) the ABCDE/FRAS tag dictionaries used in each system. Some raw industrial source code cannot be released due to confidentiality constraints; for such materials, index snapshots and derived evaluation records are provided.

\section*{Competing Interests}

The authors have filed patent applications related to the AOCI protocol with the China National Intellectual Property Administration. The AOCI Platform is developed and maintained by Xingyun Zhixue (Beijing) Technology Co., Ltd., where some of the authors hold positions.

\begin{acks}
This work was supported by the National Natural Science Foundation of China Key Program \textit{Research on the Management Theory and Policy of New Media Development} (Grant No.~71633001). The authors also thank the AI Application and Innovation Lab, School of New Media, Peking University.
\end{acks}

\section*{Use of Generative AI}

The authors used large language models (including Claude and GPT-series models) to assist with language polishing and copy-editing. All research ideas, protocol design, system implementation, experimental design, data collection, and analysis are the authors' own original work.

\clearpage

\bibliographystyle{ACM-Reference-Format}
\bibliography{ref}

\end{document}